\documentclass[aps]{revtex4}

\usepackage{graphicx}
\usepackage{psfrag}

\usepackage{bm}

\begin{document}

\title{\bf {\LARGE How Time Works in Quantum Systems:} \\
    {\Large Overview of time ordering and time correlation
    in weakly perturbed atomic collisions
    and in strongly perturbed qubits} \vskip 0.17in}

\author{J. H. McGuire$^1$, L. Kaplan$^1$, \\
    Kh. Kh. Shakov$^1$, A. Chalastaras$^1$, A. M. Smith$^1$, \\
   A. Godunov$^2$, H. Schmidt-B\"{o}cking$^3$,  \\
   and D. Uskov$^4$ }

\affiliation{ \vskip 0.1in
$^1$ Department of Physics, Tulane University, New Orleans, LA 70118, USA \\
$^2$ Department of Physics, Old Dominion University, Norfolk, VA 23529, USA \\
$^3$ Institut f\"ur Kernphysik, Universit\"at Frankfurt, 60486 Frankfurt,
Germany \\
$^4$ Department of Physics, Louisiana State University, Baton Rouge, LA 70803,
USA \vskip 0.17in}

\begin{abstract}
Time ordering may be defined by first defining the limit of no time ordering
(NTO) in terms of a time average of an external interaction, $V(t)$.
Previously, time correlation was defined in terms of a similar limit called the
independent time approximation (ITA).  Experimental evidence for time
correlation has not yet been distinguished from experimental evidence for time
ordering.
\end{abstract}

\maketitle

\vskip 3em

\newpage



\newpage

\section{Introduction}

\subsection{Space and time}

Both space and time are parameters used to mathematically describe
observable properties of physical objects or systems of objects.  In
some ways space and time are similar, both intuitively and
mathematically, since they are the most basic coordinates used to
describe the dynamics of physical systems.  It is often convenient
and conventional to use $t$ and $\vec r$ to specify when and where
an object is, even in quantum systems where the precision (or
locality) of these coordinates is limited by the uncertainty
principle or obscured by entanglement.  Nevertheless, space and time
differ fundamentally (even neglecting the second law of
thermodynamics, which defines a direction in time but no preferred
direction or arrow in space).  Because there is no counterpart of
temporal causality in a space-like context, an object can repeatedly
return to any spatial location, but it cannot return or jump ahead
temporarily in time.  Much has been written about spatial properties
of multi-particle systems, but less about time, even though
three-dimensional space would seem much harder to deal with than
one-dimensional time.  Time, unlike space, seems somehow enigmatic.
Newton refers to time as ``like a river flowing." Einstein calls
time ``that which a clock measures."  And Feynman refers to time as
the space between events, i.e. that which keeps {\LARGE
EVERYthing}{\Large FROMhappen}{ing}{\small ALLat}{\tiny ONCE}.  One
purpose of this paper is to see what happens when we try to carry an
idea formulated in a spatial context into a time context.  That idea
is correlation (associated with non-randomness, or quantum
entanglement).  Time correlation~\cite{mc01,ita} has been directly
related to time ordering, a causality condition implied by the
time-dependent Schr\"{o}dinger equation that constrains sequential
interactions to occur in order of increasing time.

\subsection{N-body problem}

It is well known that the N-body problem, namely determining the
quantum evolution of $N$ correlated or interacting bodies, is
exponentially difficult.  Kohn~\cite{kohn} specifically estimates
that the amount of computer storage capacity needed to solve an
N-body problem scales as $e^{3N}$, where 3 is simply a coefficient
based on experience.  Combining Kohn's estimate with Moore's law,
which estimates, again based on experience, a doubling of computer
capacity every 1.5 years, one can easily show that 6.5 years are
required for computer capacity to grow enough to accommodate one
additional body.  Thus, increasing capacity from, say helium (N = 3
ignoring nuclear and subnuclear structure) to carbon (N = 7) is
estimated to require about 25 years, progressing to water (N = 11
ignoring the inner shell) requires about 50 years, and DNA thousands
of times the age of the universe.  The N-body problem is difficult.
On the other hand, since much of physics, chemistry, biology,
materials science, nanotechnology, and quantum computing involves
multiple particles, the problem is significant.  In quantum
computing, for example, much has been understood about individual
qubits (regarded here as single ``bodies"), but much less is
understood about the coupled networks of qubits needed to build a
quantum computer.  Evidently, there is a need for sensible,
well-defined approximations to the N-body problem.  Perhaps the most
widely used of these is the independent particle approximation
(IPA), where the N-body problem is dramatically reduced in
complexity to a problem of N independent bodies (or particles,
quasi-particles, qubits, \ldots).

\subsection{Correlation}

It is sometimes argued that a theory is not well defined until a reliable
method is developed to calculate corrections to the theory.  In the case of the
IPA, the corrections are called correlations and represent interconnections
among the positions of the various particles.  Correlation implies complexity:
a system of independent (i.e. uncorrelated) particles is less complex than a
system in which each particle's behavior depends on the behavior of every other
particle.  From one viewpoint, correlation is a key to understanding how to
make complex systems from simple ones.  From the opposite point of view, in a
very complicated system correlation can be a pathway to see order through a
landscape of chaos.

Our approach is to focus on the first viewpoint -- building complex
systems from simple ones.  If the correlations are sufficiently
small, then the IPA is reasonably accurate.  Correlation,
one method to approach the N-body problem, is defined as the
difference between a full solution and the IPA.  Thus, the IPA limit
of no correlation is used to define correlation itself.  While this
approach appears a bit awkward at first, it is conventional in the
study of non-random processes and quantum entanglement as well.  For
example, in the limit of no entanglement the N-body wave function
may be written as a simple product of independent particle
solutions, and entanglement may be defined as the deviation from
this limit.

In most applications, correlations arise from spatial interactions
between particles.  In N-electron atoms, for example, electron
correlation is generated by the $1/r_{ij}$ Coulomb interactions
between electrons.  In this context, the IPA is also known as the
independent electron approximation (IEA) and is defined by replacing
these complicating correlation interactions by a sum of
single-electron mean-field interactions.  The resulting Hamiltonian
is a sum of single-electron terms, even in the case of dynamic
atomic systems~\cite{mcbook}, and the resulting wave function is a
product of single-electron wave functions.  One common method for
analyzing deviations from the IEA is many-body perturbation theory
(MBPT), used here and in other fields as well to describe
correlation at various orders in a given perturbation.

One of the primary questions motivating the work described here is,
``what happens if one tries to define an independent time
approximation (ITA), where one eliminates correlations in time
rather than the conventional correlations in space?"  In such an
approximation, multiple times may be used for multiple
particles~\cite{mg03}.  This is simpler than using a single time for
all particles, just as the use of N independent particle positions
is simpler that trying to solve a single, but complicated, N-body
problem.  Time correlation may then be defined as a deviation from
the ITA in analogy to the definition of spatial correlation by
reference to the IPA.

\subsection{Correlation in time and time ordering}

In this paper we first review early efforts to define the ITA, in
which a key step is to remove time ordering between fields acting on
different particles.  This is closely related to the approximation
of no time ordering (NTO), where all time ordering is removed for
fields acting either on different particles or on the same particle.
These early efforts to define and study the ITA were based on a
perturbation expansion of the external interaction $V(t)$.  Since
time correlation or time ordering first appear at second order in
$V(t)$, the observable effects are small and the calculations
difficult.  Next we review more recent investigations of time
ordering and the NTO approximation for qubits in non-perturbative
external fields.  Therefore, the organization of this paper is
unusual.  The harder problem of time correlation is addressed first,
and the related but more tractable problem of time ordering is
discussed second.  We follow this unusual order because it reflects
what has happened in recent years and also because the newer work on
time ordering raises some intriguing questions and challenges.  In
particular we emphasize that experimental evidence for time
correlation has not yet been distinguished from evidence for time
ordering.  That is, no difference between the ITA and the NTO
approximations has yet been observed in any experiment, although
recent experiments have been progressing in this direction, as we
illustrate below.


\section{The N-body problem}

In this section we consider solutions to the N-body problem described by a
Hamiltonian
\begin{eqnarray}
\label{H}
  \hat H = \hat H_0 + \hat V_S(t)  \,,
\end{eqnarray}
where there may be many terms in both the unperturbed $\hat H_0$,
which we assume to be solvable, and in $\hat V_S(t)$, which we
regard as an external time-dependent interaction (written in the
Schr\"odinger representation).  Where appropriate, $\hat V_S(t)$ may
be treated as a classical external field.  In the case of atomic
collisions, this corresponds to the widely used semiclassical
approximation (SCA) where the projectile is regarded as a classical
particle if it is a proton or electron and a classical wave if it is
a photon.

The N-body problem may be solved in various representations,
depending on how $\hat H$ is separated into $\hat H_0$ and $\hat
V_S(t)$, ranging from the Schr\"{o}dinger representation where
effectively $\hat V_S(t) = \hat H$ and $\hat H_0=0$ to the
Heisenberg representation where $\hat H_0=\hat H$ and $\hat
V_S(t)=0$.  In the SCA, singularities can occur~\cite{mcbook} in the
first order amplitudes when the Schr\"{o}dinger representation is
used.  Except where otherwise specified we shall use an appropriate
intermediate representation that takes maximum advantage of known
solutions for $\hat H_0$, and in which $\hat V(t)=e^{i\hat H_0
t}\hat V_S(t)e^{-i\hat H_0 t}$ causes transitions between
eigenstates of $\hat H_0$.  We work in atomic units, in which $e^2 =
m_e = \hbar = 1$.

\subsection{Formulation of the N-body problem in the time domain}

We seek solutions $\Psi(r_1, \ldots, r_N;t)$ to the N-body problem described by
Eq.~(\ref{H}).  It is both conventional and sensible to separate the influence
of the external interaction $\hat V(t)$ from the initial state at some time
$t_0$ before the external fields are applied to the atomic system, namely,
\begin{eqnarray}
\label{Nbody}
 \Psi(r_1, \ldots, r_N;t) =  \hat U(t,t_0) \Psi(r_1, \ldots, r_N;t_0) \,.
\end{eqnarray}
Here all dynamics are contained in the time evolution operator (or Green's
function) $\hat U(t,t_0)$.

It is easily shown that in the intermediate representation $\hat U(t,t_0)$
satisfies
\begin{eqnarray}
\label{deU}
 i \frac{d}{dt} \hat U(t,t_0) = \hat V(t) \, \hat U(t,t_0) \,.
\end{eqnarray}
The solution for $\hat U(t,t_0)$, which may be verified by insertion
into Eq.~(\ref{deU}), is
\begin{eqnarray}
\label{U} \hat U(t,t_0) &=& 1 -i \int_{t_0}^{t} \hat V(t_1) \, dt_1
+ (-i)^2 \int_{t_0}^{t} \hat V(t_1) \, dt_1 \int_{t_0}^{t_1} \hat
V(t_2) \, dt_2 \nonumber \\ &+& \cdots \,+ (-i)^n \int_{t_0}^{t}
\hat V(t_1) \, dt_1 \int_{t_0}^{t_1} \hat V(t_2) \, dt_2 \, \cdots
\int_{t_0}^{t_{n-1}} \hat V(t_n) \, dt_n + \,\cdots \nonumber \\
&\equiv&  T \sum_n \frac{(-i)^n}{n!} \left[ \int_{t_0}^{t} \hat
V(t') \, dt' \right]^n
    = T e^{-i \int_{t_0}^t \hat V(t') \, dt'} \,.
\end{eqnarray}
Here $T$ is the Dyson time-ordering operator, which arranges the interactions
$\hat V(t')$ in order of increasing time, similar to the requirements of
causality.  The time ordering operator is central to the discussion in this
paper, as it relates both to observable time ordering effects and to time
correlation.  The key idea is that $T \, \hat V(t_1) \hat V(t_2) = \hat V(t_2)
\hat V(t_1)$ if $t_2 > t_1$ and $\hat V(t_1) \hat V(t_2)$ otherwise. {\it The
no time ordering (NTO) approximation is the approximation in which $T \to 1$,
and the constraint of time ordering is not enforced.} In this limit all time
sequences of the $\hat V(t')$ are equally weighted, so that
\begin{equation}
\hat U(t,t_0) \to \hat U_{\rm NTO}(t,t_0)= e^{-i \int_{t_0}^t \hat
V(t') \, dt'}
\end{equation}
in Eq.~(\ref{U}).

\subsection{Independent particle approximation (IPA)}

As noted above, the N-body problem is notoriously difficult to
solve.  A particularly useful approximation is the independent
particle approximation (IPA).  For an atom with N electrons, $\hat
H_0 = \sum_j ( -\nabla_j^2/2 - Z/r_j + \sum_{i< j} 1/r_{ij})$, where
$Z$ is the nuclear charge and $\vec r_j$ are the electron positions
in a coordinate system centered on the nucleus, and $\hat V_S(t) =
-\sum_j Z_p/|\vec{R}(t) - \vec{r}_j|$, where $Z_p$ and $\vec R(t)$
are the charge and position of the projectile.  If $\sum_{i<j}
1/r_{ij}$ is approximated by a mean field $\sum_j \hat v(\vec r_j)$,
then $\hat H = \hat H_0 + \hat V_S(t)$ is reduced to a sum of
single-particle terms that may be solved using separation of
variables.  One obtains the independent particle approximation (IPA)
to the exact N-particle wave function, namely,
\begin{eqnarray}
\label{IPA}
 \Psi(r_1, \ldots,r_N;t) \simeq \prod_j\psi_j(r_j,t) = \prod_j \hat
 U_j(t,t_0) \,
\psi_j(r_j,t_0) \ \ \ \ ({\rm IPA}) \,,
\end{eqnarray}
where each $\psi_j(r_j,t)$ is a single-particle wave function.  We
note here that in the IPA approximation, different times may be used
for different particles, if so desired, since the particles are
independent in both space and time.  Whatever happens to one of the
particles does not influence what happens to any of the others,
although time ordering is retained within the evolution of each
independent particle.

\subsection{Independent time approximation (ITA)}

In order to address the question, ``what happens if one tries to
define an independent time approximation (ITA), where one explores
correlations in time rather than the conventional correlations in
space?", a list of comparisons between the ITA and IPA was
developed~\cite{ita}.  An updated comparison is summarized in
Table~\ref{table1}.

\begin{table}
\caption{Comparison of correlation in space and time.}
\begin{tabular}{||@{}lll||}  \hline\hline
     & Spatial correlation & Temporal correlation \\
\hline\hline
\hspace{0.3em} Cause:   & $\hat v_{ij} = 1/r_{ij}$ &
$T$ and $\hat
V(t)$  \\ & spatially varying internal             & time ordering of    \\
         & Coulomb interactions                  & external interactions \\

\hline & & \\[-10pt] \hspace{0.3em} Origin: & $\hat H_0 = \sum_j \hat H_{0j}
+ \sum_{i<j}\hat
v_{ij}\;\;\;\;\;\;$ & $ i {d\over dt}{\hat U}  =  \hat V(t)\, \hat U $
\\[2pt]

\hline \hspace{0.3em} Uncorrelated limit: $\;\;$&  IPA    &   ITA   \\
\hline & &\\[-10pt] \hspace{0.3em} Product form: & $\Psi(r_1,\ldots,r_N) \to \prod_j
\psi_j(r_j)$
         &  $\Psi(t_1,\ldots,t_N) \to  \sum_k c_k \prod_j \psi^{(k)}_j(t_j)$
\\[2pt] \hline & & \\[-10pt] \hspace{0.3em} No fluctuations: & $\hat v_{\rm cor}  = \hat v_{ij} -
\hat v_{\rm av} \to  0 $ & $T_{\rm cor} =  T - T_{\rm av} \to 0
\;\;$ or $\;\;\delta \hat V(t)=\hat V(t)-\overline{\hat V} \to 0$ \\[2pt] \hline & & \\[-10pt] \hspace{0.3em} Average value: & $\hat v \to \hat
v_{\rm av} = \hat v_{\rm mean \ field} $ &  $T \to T_{\rm av} =
1\;\;$ or $\;\;\hat V(t) \to \overline{\hat V}$
\\[2pt]  \hline\hline
\end{tabular}
\label{table1}
\end{table}

There are similarities between the temporal independent time
approximation and the spatial independent particle approximation, as
seen in Table~\ref{table1}.  Time and space correlation can each be
defined as a deviation from an uncorrelated limit, where the
uncorrelated limit is given by a product form.  Electron identity,
which has been ignored here for simplicity of presentation, may be
restored by antisymmetrizing the uncorrelated single-electron wave
functions. The uncorrelated limit may also be described by an
average of the appropriate correlation operator, as indicated in
Table~\ref{table1}.  Correlation may then be defined in terms of
fluctuations away from the average, as is done in statistical
mechanics~\cite{Balescu}.  In both the spatial and temporal cases,
the average term may form the basis for useful approximate
calculations.

There are also notable differences~\cite{time} between temporal and
spacial correlation, as detailed in Table~\ref{table1}.  While
correlation in space arises in the asymptotic target Hamiltonian
$\hat H_0$, and affects both the asymptotic initial wave function
$\Psi(t_0)$ and the evolution operator $\hat U(t,t_0)$, correlation
in time occurs only in the time evolution operator $\hat U(t,t_0)$.
Correlation in space comes from $1/r_{ij}$ inter-electron
interactions within the target.  In the IPA, phase coherence
and time correlation between electrons are both lost,
as seen, for example, by noting that matrix elements of
$\hat{V}$ in Eq.~(\ref{U2}) may be complex and that the
time order is significant except when all the $\hat{V}$
go to a common $\overline{\hat V}$.
Time correlation
arises from time ordering of the external interaction $\hat V(t)$
acting on different particles.  The ITA is intimately related to the
NTO since $T \to T_{\rm av}=1$ in both cases. However, in the ITA $T
\to 1$ is applied only to the cross terms affecting different
particles, while time ordering for each individual particle is
retained.  In the ITA, each particle evolves independently in time,
although the initial state may be spatially correlated.  Thus, the
initial state $\sum_k c_k \prod_j \psi^{(k)}_j(\vec r_j,t_0)$
evolves to $\sum_k c_k \prod_j \psi^{(k)}_j(\vec r_j,t_j)$, where
the state of each particle may be evolved using its own independent
time $t_j$.  Removing some or all of the time ordering terms is
straightforward in practice since the $T_{\rm cor}$ terms are easily
identified, at least at second order in perturbation
theory~\cite{mc01}.


\section{Atomic collisions}

In this section we review time ordering and time correlation using
perturbation theory for the interaction of charged particles with
helium~\cite{mc01,ita}.  From Eq.~(\ref{U}) we see that the leading
effect due to time ordering arises at second order in the $\hat
V(t)$ expansion.  Thus the system has an infinite number of states
but the external interaction $\hat V(t)$ only acts on the system
twice.  In Sec.~\ref{secqubits}, we consider a strongly perturbed
qubit, a two-state system that interacts with the external field
$\hat V(t)$ an infinite number of times.

Through second order in $\hat V(t)$, the time evolution operator $\hat
U(t,t_0)$ is given by
\begin{eqnarray}
\label{U2}
 \hat U(t,t_0) &=& T e^{-i \int_{t_0}^{t} \hat V(t') \, dt'}
\simeq 1 -i \int_{t_0}^{t} \hat V(t_1) \, dt_1
- \int_{t_0}^{t} \hat V(t_1) \, dt_1  \int_{t_0}^{t_1} \hat V(t_2) \, dt_2
  \nonumber \\
    &\equiv&  1 -i \int_{t_0}^{t} \hat V(t_1) dt_1
- T \frac{1}{2!} \int_{t_0}^{t} \hat V(t_1) \, dt_1  \int_{t_0}^{t}
  \hat V(t_2) \, dt_2 \,. \end{eqnarray}
This may also be obtained by integrating Eq.~(\ref{deU}) through
second order.  Note that as $T \to 1$ the order in which the
interactions act may be interchanged.  It is the difference between
the limit as $T \to 1$ and the full result that defines the effects
of time ordering.  In the case of weak correlation considered here,
the NTO limit of $T \to 1$ will coincidentally yield the ITA as
well.  The trick now is to separate $T$ from $T_{\rm av} = 1$.

\subsection{Primary results}

\subsubsection{Various pathways to ITA}
\label{secpathways}

   To separate the time ordering effects from the non-time ordering
(NTO) effects, it is useful to write,
\begin{eqnarray}
\label{T-Tav}
  T = T_{\rm av} + (T - T_{\rm av})
\end{eqnarray}
where $T_{\rm av}$ yields the NTO approximation and $T - T_{\rm av}$
yields the effects of time ordering.  This decomposition of $T$ into
an average part plus fluctuations is central for this paper.  Note
that $T_{\rm av}=1$ is required to satisfy the initial condition
$\hat U(t_0,t_0) = 1$.

One conceptual pathway to the ITA proceeds by analogy with the NTO
limit.  In the ITA, time ordering is enforced among all potentials
acting on an individual particle (e.g. for $\hat V_i(t_1)\hat
V_i(t_2)$ at second order), but ignored for different particles in
cross terms such as $\hat V_i(t_1)\hat V_j(t_2)$ when $i \ne j$.
This distinguishes the ITA from the NTO approximation where time
ordering is removed for all terms.

Another clever conceptual pathway to time ordering and the ITA, noticed by
Godunov~\cite{ita}, is via use of commutator relations.  Consider the identity,
which we call the Godunov identity~\cite{gc},
\begin{equation}
\label{Crel}
\hat V(t_1) \hat V(t_2) = \frac{1}{2}\left (\hat V(t_1) \hat V(t_2) + \hat
V(t_2) \hat V(t_1)\right ) +
\frac{1}{2} [\hat V(t_1),\hat V(t_2)] \,.
\end{equation}
In the case of atomic collisions with helium, the projectile
interacts with both electrons so that $\hat V(t') = \hat V_1(t') +
\hat V_2(t')$.   Now $[\hat V(t_1),\hat V(t_2)]$ contains both
$[\hat V_i(t_1),\hat V_i(t_2)]$ and $[\hat V_i(t_1),\hat V_j(t_2)]$
($i \neq j$) terms.  In the NTO approximation, all commutator terms
are eliminated, i.e. $[\hat V(t_1),\hat V(t_2)] \to 0$ and $\hat
V(t_1) \hat V(t_2) \to \frac{1}{2}\left(\hat V(t_1) \hat V(t_2) +
\hat V(t_2) \hat V(t_1)\right)$ so that the time evolution operator
at second order is given by the average value of both possible time
sequences.  In the ITA, only cross commutator terms between fields
acting on different particles are neglected, i.e. $[\hat
V_i(t_1),\hat V_j(t_2)] \to 0$ and $\hat V_i(t_1) \hat V_j(t_2) \to
\frac{1}{2}\left(\hat V_i(t_1) \hat V_j(t_2) + \hat V_j(t_2) \hat
V_i(t_1)\right)$ only for $i \ne j$.

A third pathway to the ITA is through elimination of
off-energy-shell effects.  The effect of the Dyson time ordering
operator  at second order may be expressed as $T \, \hat V(t_1) \hat
V(t_2) = \Theta(t_1 - t_2) \hat V(t_1) \hat V(t_2) +(1
\leftrightarrow 2)$, and thus
\begin{equation}
\label{thetafun} {1 \over 2!} \, T  \int_{-\infty}^{+\infty}
\int_{-\infty}^{+\infty}  \hat V(t_1) \hat V(t_2) \, dt_1 \, dt_2 =
\int_{-\infty}^{+\infty} \int_{-\infty}^{+\infty} \Theta(t_1 - t_2)
\hat V(t_1) \hat V(t_2) \, dt_1 \, dt_2 \,.
\end{equation}
Taking $T \to 1$ is equivalent to replacing $\Theta(t_1-t_2)$ by the constant
$1/2$.  The Fourier transform of the Heavyside theta function is well known,
namely
\begin{eqnarray}
\label{FTthetafun}
   \int_{-\infty}^{+\infty} e^{i E (t_1-t_2)}
 \Theta(t_1 - t_2) \, d(t_1-t_2) =
\pi \delta(E) + i P_v \frac{1}{E} \,.
\end{eqnarray}
The principal value term accesses off-energy-shell states, which
violate energy conservation during the short collision time in a
manner consistent with the uncertainty relation.  Since taking $T
\to 1$ gives the $\pi\delta(E)$ term in Eq.~(\ref{FTthetafun}), it
is the principal value term that carries the effects of time
ordering.  Ignoring the off-shell contribution (which is
conveniently $\pi/2$ out of phase with the time-averaged term)
whenever the fields $\hat V(t_1)$ and $\hat V(t_2)$ act on different
particles yields the ITA.

There is yet a fourth pathway to the NTO or ITA.  For potentials
that are time-independent in the intermediate representation, $[\hat
V(t_1),\hat V(t_2)] \to 0$ and the NTO approximation is exact.  In
general, the NTO approximation may be obtained by replacing the true
external potential $\hat V(t)$ with its time average over the
duration of the interaction, i.e. $\displaystyle \hat V(t') \to
\overline{\hat V}={1 \over t-t_0} \int_{t_0}^t \hat V(t') \, dt'$,
so that the exponent $\int \hat V(t') \, dt'$ in the time-ordered
exponential of Eq.~(\ref{U}) is replaced by $\overline{\hat V}\cdot
(t-t_0)$.  The replacement of the external time-dependent
interaction by its time average is analogous to the replacement of
the true inter-particle interaction by its mean-field value in the
IPA, which involves averaging over the positions of all but one of
the particles.

We note that the limit of constant potential $\hat V(t)$ in which
the NTO approximation becomes exact is distinct from the adiabatic
limit, in which the potential merely changes slowly with time.
Furthermore, the content of the NTO or ITA approximation depends on
the representation used, as we will see explicitly in
Sec.~\ref{secqubits}.  Thus, for a given decomposition $\hat H=\hat
H_0+\hat V_S(t)$, it is the interaction-representation potential
$\hat V(t)=e^{i \hat H_0 t}V_S(t)e^{-i\hat H_0 t}$ that must be
constant for the NTO or ITA approximation to be exact.  In the
Heisenberg representation, $\hat V(t)=0$ by construction, and the
NTO or ITA is always trivially exact.  To summarize, pathways to the
ITA approximation in this section include:

\begin{enumerate}

\item $T \to 1$,

\item $[\hat V_i(t_1),\hat V_j(t_2)] \to 0$,

\item $ P_v \frac{1}{E} \to 0$,

\item $\hat V(t) \to \overline{\hat V}$\,.
\end{enumerate}
The difference between the ITA and NTO approximations is that in the
second item above, the commutator disappears for all terms in the
NTO but only for the cross terms in the ITA.

\subsubsection{Economy of NTO}

A variety of second-order calculations with and without time
ordering have been done by Godunov and collaborators in recent
years~\cite{gm01,god05}.  Most of these calculations have studied
two-electron transitions in atoms caused by high-velocity collisions
with protons or electrons, where second Born methods are applicable.
The advantage of looking at two-electron transitions is that second
order terms are often dominant, since both elastic scattering and
single-electron transitions, corresponding to the first two terms in
Eq.~(\ref{U2}), are experimentally eliminated.  A disadvantage is
that the resulting cross sections are quite small.  The Godunov code
is remarkable in that the off-shell terms can be computed exactly at
second order in perturbation theory, in contrast with most other
existing calculations, which use closure approximations to avoid
this relatively difficult calculation.  Dropping the difficult but
interesting off-shell terms yields a result without any time
ordering.  Comparing results with and without inclusion of off-shell
terms then yields the time ordering or time correlation effect.  The
algebra required for the off-shell terms is relatively tedious, as
reflected by the off-shell calculation typically requiring several
hundred times more computer time.  Thus the NTO (and possibly the
ITA) represents a substantial reduction of computer time and
algebraic effort.  When valid, this approximation can therefore be
used to attack problems harder than those that require the full
off-shell terms.  Unfortunately, at this point no simple, physically transparent
criterion is known to us that determine when the NTO or ITA approximation is valid.

\subsection{Experimental evidence}

There have been ten or so experiments that show some evidence for
time ordering and time correlation effects for two-electron
transitions in high velocity collisions.  Perhaps the most dramatic
evidence is the factor of two difference found~\cite{aarhus} in the
double ionization of helium in collisions with protons and
antiprotons at several MeV.  There seems to be agreement that this
difference is due to time ordering and time correlation, but no
definitive theoretical studies have yet been done.  Some
studies~\cite{god97} of Auger profiles also support the need for
time correlation and time ordering.  The clearest direct comparison
of experimental data with time-ordered and non-time-ordered
theoretical calculations is for polarization of light emitted after
excitation-ionization of helium by proton impact~\cite{polarize}.

\begin{figure}[h]
\begin{center}
\includegraphics[keepaspectratio=true,width=0.50\textwidth]{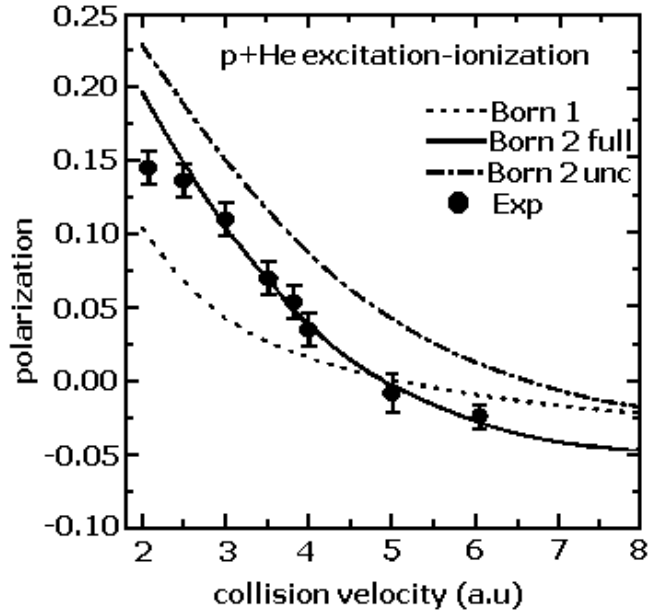}
\caption{\footnotesize{\label{f1} Calculations with and without time
ordering between electrons are compared to experimental data.  Here
polarized light is emitted from helium following $1s \to 2p$
excitation of one electron accompanied by ionization of the second
electron~\cite{polarize}.  The polarization fraction is plotted as a
function of the velocity of the incident proton.  The first-order
calculation (Born 1) has no time ordering. The second-order
calculation is shown both with time ordering (Born 2 full) and in
the no time ordering (NTO) approximation (Born 2 unc).  As explained
in the text, the NTO approximation is the same as the
independent time approximation (ITA) at this order since the
interaction is weak.}}
\end{center}
\end{figure}

In this study, time ordering or time correlation has a $20\%$ effect
on the predicted polarization.  The experiment, with $5\%$ errors,
is in excellent agreement with Godunov's calculations, except at the
lowest energy where perturbation theory is expected to break down,
and clearly shows the importance of time ordering or time
correlation effects.  There is a need for more calculations by
different authors to confirm these effects of time correlation and
time ordering.

In all of these two-electron transition experiments, one may do a simultaneous
expansion in the external interaction $\hat V(t)$ and the correlation
interaction $\hat v$.  Time correlation effects are associated with commutators
of the form $[\hat V_i(t_1),\hat V_j(t_2)]$, where $i$ and $j$ label two
different electrons, and are second order in $\hat V$.  In addition, there are
time ordering effects not related to time correlation: these are associated
with commutators $[\hat V_i(t_1),\hat V_i(t_2)]$ and enter at second order in
$\hat V$ and first order in $\hat v$ (since the commutator vanishes for $\hat
v=0$).  Thus the difference between time ordering and time correlation is a
third order effect, smaller by at least a factor of 10 than either the time
ordering effect or the time correlation effect taken individually.  The
experiments show direct evidence for time correlation but do not distinguish
between time correlation and time ordering.

There is also one clear and definitive study by the group of
Thomas~\cite{zhao} that shows direct evidence for time ordering in
atoms interacting with a time-varying magnetic field.


\section{Qubits}
\label{secqubits}

A qubit is a very simple two-state (e.g. on and off) quantum system
whose state population may be changed by an external potential $\hat
V(t)$, in analogy with the way in which an atomic state may be
changed by the Coulomb potential $\hat V(t)$ of the projectile in an
atomic collision.  Qubits are building blocks for the complex
interconnected N-qubit systems that can be used for quantum
computation and quantum information.  The extension from a single
qubit to a system of N interconnected qubits is analogous to the
extension from a one-electron atom to a correlated N-electron atom.
The interaction of qubits with each other and with their environment
still have to be dealt with before a quantum computer becomes a
reality.

In this section we consider time ordering in a single qubit,
analogous to time ordering in scattering from atomic hydrogen.  To
better understand time ordering effects, we work in the time domain
rather than the more common energy (or frequency) domain.  The
advantages of working with qubits include the possibility of easily
handling non-perturbative external potentials and the existence in
some cases of analytic solutions, so that numerical calculations can
be avoided.  This yields new ways to think about time ordering
mathematically and physically.  Specific effects due to time
ordering in a simply pulsed qubit are shown, for example, in
Fig.~\ref{figdifference} below.  The idea is to extend the above
analysis of time ordering for weakly perturbed atomic collisions to
the case of strongly perturbed qubits.  In order to make use of
transparent analytic expressions, we spend some effort discussing
the time evolution of qubits pulsed sharply in time, commonly
referred to as ``kicked" qubits.  With the exception of the last
brief subsection, there is little emphasis here on time correlation
between qubits, simply because little work has been done on this
problem to our knowledge.  So our emphasis in this section is almost
entirely on time ordering.  One new feature here will be to explore
what happens to time ordering under change of representation --
specifically when we change from the Schr\"{o}dinger to the
intermediate representation.

\subsection{Overview of single qubits}

The qubit wave function $\psi(t)$ is a linear superposition of the
``on" and ``off" states, namely

\begin{eqnarray}
\label{psi} \Psi(t) = a_1(t) \left[
\begin{array}{c} 1 \\ 0 \end{array} \right] + a_2(t) \left[
\begin{array}{c} 0 \\ 1 \end{array} \right] \,.
\end{eqnarray}

Population can be transferred from one state to the other by
applying an external potential $\hat V(t)$, which can have the form
of a single pulse characterized by a time duration, $\tau$.  The
full Hamiltonian can be written in terms of the Pauli spin matrices,

\begin{eqnarray}
 \label{genh0} \hat H(t) &=& \hat H_0 + \hat V_S(t) \nonumber\\
           &=& \left[ \begin{array}{cc} -\Delta E/2 & 0 \\ 0 & +\Delta E/2
\end{array} \right] + \left[ \begin{array}{ccc} 0 & V(t) \\ V(t) &
0 \end{array} \right]  \\
 &=& -{\Delta E \over 2} \sigma_z + V(t) \sigma_x. \nonumber
\end{eqnarray}
The time dependence of the system can be described by the time
evolution operator $\hat{U}(t, t_0)$.  Allowing $t_0 = 0$ to
simplify the notation, we can write $\psi(t)=\hat{U}(t)\psi(0)$,
where

\begin{eqnarray}
\label{uoperator} \hat{U}(t) = \left [ \begin{array}{cc}
  U_{11}(t)  & U_{12}(t)  \\
  U_{21}(t)  & U_{22}(t)
\end{array} \right] = Te^{-i \int_{0}^{t} \hat{V}(t') dt'}
\end{eqnarray}
Here $\hat V(t')=e^{i\hat H_0 t'}V_S(t')e^{-i\hat H_0 t'}$ is the
interaction potential in the intermediate representation, and $\hat
U(t)$ is the formal solution to the differential equation
(\ref{deU}).  In the Schr\"odinger representation, $\hat V(t')$ is
replaced by $\hat H(t')=\hat H_0+\hat V_S(t')$.

The time evolution of a qubit depends on the energy splitting
$\Delta E$ and the time dependence $V(t)$ of the external potential.
Depending on the complexity of $V(t)$, $\hat U(t)$ may or may not
have a simple analytic form.  In the latter case, numerical
solutions of the coupled differential equations (\ref{deU}) may
obscure information about these quantum systems.  Analytic
solutions, when available, are more convenient and easy to analyze.

\subsection{Simply pulsed qubits}

Here we consider simply pulsed qubits -- that is, qubits subject to
an external field $V(t)$ that has a finite duration in time and a
sensibly simple shape, such as a simple rectangular or Gaussian
pulse. Such pulses are convenient for studying qubits in the time
domain. In some cases they also lead to convenient analytic
solutions, even for strong fields.

\subsubsection{Qubit map}

\begin{figure}
\begin{center}
\psfrag{xy}{$0$}
 \psfrag{x1}{$2 \pi$}
 \psfrag{x2}{$\infty$}
 \psfrag{x3}[c][c][1.5][0]{${\int_0^\infty {V(t')}
 dt'}$}
 \psfrag{y1}[c][c][1][90]{$2 \pi$}
 \psfrag{y2}[c][c][1][0]{$\infty$}
 \psfrag{y3}[c][c][1.5][90]{$\frac{\Delta E \,\tau}{2}$}
 \psfrag{z1}[l][l][1][0]{$\tau \rightarrow 0$ Fast}
 \psfrag{z2}[c][c][1][0]{$\tau \rightarrow \infty$ Slow}
\includegraphics[keepaspectratio=true,width=0.65\textwidth]{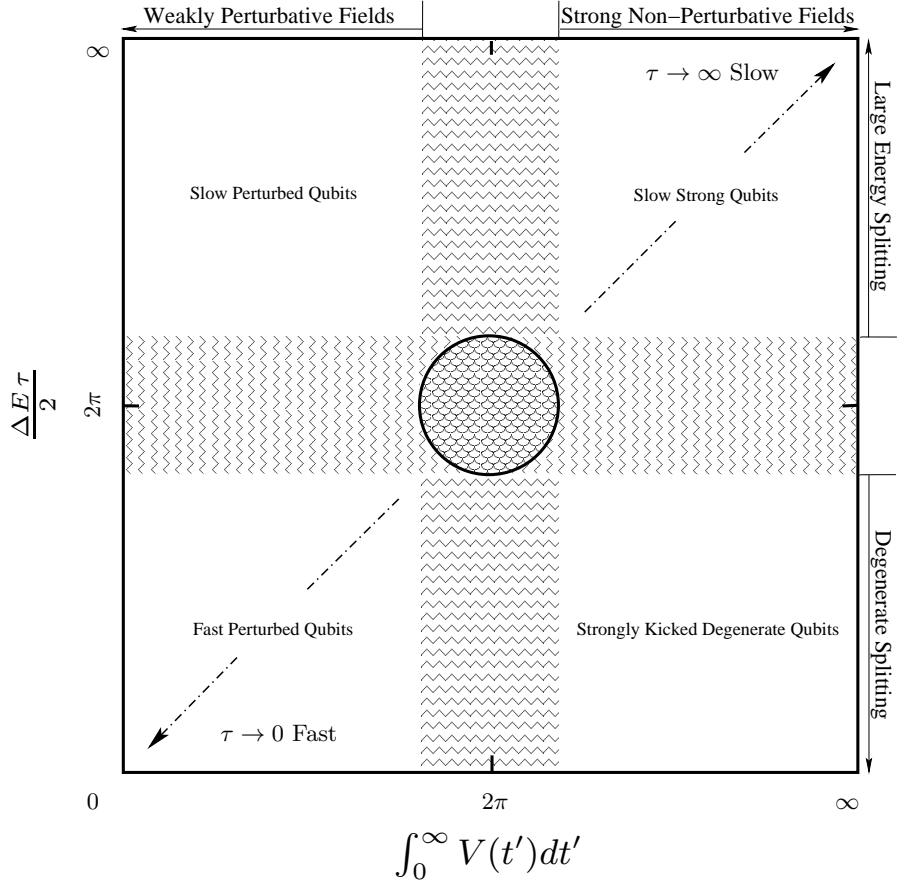}
 \caption{\footnotesize{\label{f2} Qubit map for qubits
interacting with simply pulsed finite external potentials.  Here
$\tau$ is the duration of the pulse, $\Delta E$ is the energy
difference between the two states of the qubit, and $V(t)$ is the
external potential.  When the total phase associated with the
external potential is small, i.e. $\int_0^\infty {V(t')} \,dt' \ll
2\pi$, then the expression for the time evolution operator may be
expanded in powers of $V$ using $e^{-iV(t')\,dt'} \approx
1-iV(t')\,dt'$ and only the first few terms retained.  This
corresponds to standard perturbation theory for either quickly
perturbed qubits where $\tau$ is small or slowly perturbed qubits
where $\tau$ is large.  Similarly, if the phase associated with the
energy splitting is small, $\Delta E\,\tau/2 \ll 2\pi$, then we can
treat the qubit as degenerate, and the solution can be expanded in
powers of $\Delta E \,\tau$.  If both phases are large, then the
adiabatic approximation generally applies.}} \label{figmap}
\end{center}
\end{figure}

A qubit map, such as that shown in Fig.~\ref{figmap}, is a tool
enabling one to visualize how the possible behavior of simply pulsed
qubits depends on the variables $\Delta E$ and $V(t)$.  In order to
make the qubit map a helpful visual tool, the map coordinates are
taken to be a dimensionless level splitting $\Delta E\,\tau/2$
(where $\tau$ is the time duration of the pulse $V(t)$) and a
dimensionless pulse strength $\int^\infty_0 V(t') \,dt'$. These two
coordinates determine the effect of the unperturbed Hamiltonian
$\hat{H_0}$ and of the external potential $\hat V(t)$ on the
evolution operator $\hat{U}(t)$.  It is useful to think of these two
variables as independent phase angles or action-like integrals.

\subsubsection{RWA solutions}

The kicked qubits we focus on below provide an alternative to the
well-established rotating wave approximation (RWA)
method~\cite{ae,shore,me} based on a single resonant transition
frequency, that may be detuned.  Our emphasis here is on kicked
(i.e. strongly time-localized) potentials because: i) the physics is
then naturally analyzed in the time domain, and ii) kicks have been
less widely explored than the RWA approach, which works well for
sharp pulses in the reciprocal frequency space.  While the RWA
method is useful for two-state systems perturbed by an external
interaction of narrow bandwidth, it fails to describe simply kicked
two-state systems where the pulse bandwidth may be very broad.

\subsubsection{Kicked qubits}
A useful approximation for simply pulsed qubits is the fast, narrow
pulse or ``kick" limit in which the width $\tau$ of the pulse goes
to zero, while the integrated strength or area under the external
potential curve
\begin{equation}
\alpha=\int_0^t V(t') \, dt'
\end{equation}
remains fixed.  Formally, the shape of a very narrow pulse of finite
total strength $\alpha$ may be expressed by a delta function:
$V(t')=\alpha \,\delta(t'-t_k)$, where $t_k$ is the time at which
the pulse is centered.  The kicked region corresponds to the lower
half of the qubit map in Fig.~\ref{figmap}.  Here the duration of
the pulse is so short that $\Delta E\,\tau/2 \ll 2\pi$, i.e. there
is not enough time for the splitting $\Delta E$ to have a
significant effect while the pulse is active.  The integrated
strength of the pulse, $\alpha$, may be either large or small in
this region.  If $\alpha$ is large, we are in the lower right
quadrant of the map, where the kicked region overlaps with the
adiabatic region.  If $\alpha$ is small, we are in the lower left
quadrant, where the kicked region overlaps with the perturbative
region.

\vskip 1em
{\bf Single kick}
\vskip 1em

For a two-state system subjected to a single kick at time $t_k$,
corresponding to $V(t') = \alpha \, \delta(t'-t_k)$, the integration
over time is trivial and the time evolution operator in
Eq.~(\ref{uoperator}) becomes
\begin{eqnarray}
\label{Usk} \hat U^k(t) &=& T \exp \left[ -i\int_0^t e^{- i \Delta
E\sigma_z t'/2} \alpha \, \delta(t'-t_k) \sigma_x e^{i \Delta E\,
\sigma_z t'/2} dt' \right]
\nonumber \\
&=& \exp \left[ -i \alpha e^{-i \Delta E \,\sigma_z t_k/2}
\sigma_x e^{i \Delta E \,\sigma_z t_k/2} \right] \nonumber \\
&=& \exp \left[ -i \alpha \pmatrix{ e^{-i \Delta E \,t_k/2} & 0 \cr
0 & e^{i \Delta E \,t_k/2}} \pmatrix{0 & 1 \cr 1 & 0} \pmatrix{ e^{i
\Delta E \,t_k/2} & 0 \cr
0 & e^{-i \Delta E \, t_k/2}} \right] \nonumber \\
&=& \exp \left[ -i \alpha \pmatrix{ 0 & e^{- i \Delta E \, t_k} \cr
e^{i \Delta E \, t_k} & 0} \right] \nonumber \\
&=& \pmatrix{ \cos\alpha & -ie^{-i \Delta E \, t_k} \sin\alpha \cr
-ie^{i \Delta E \, t_k} \sin\alpha & \cos\alpha}
\end{eqnarray}
for $t > t_k$.  The last line of Eq.~(\ref{Usk}) was obtained by
expanding the fourth line in powers of $\alpha$ using the identity
$\pmatrix{ 0 & e^{- i \Delta E \, t_k} \cr e^{i \Delta E \, t_k} & 0
}^{2n} = I$.  Equivalently, one can take advantage of the useful
identity (which we also used in the third line) $e^{i\phi\,
\vec{\sigma} \cdot \hat{u} } = \cos \phi \ \ \hat{I} + i \sin \phi \
\ \vec{\sigma} \cdot \hat{u}$, where $\hat{u}$ is an arbitrary unit
vector.  Note that $\hat U^k(t)$ is independent of the final time
$t$ in the intermediate representation.

The occupation probabilities for a kicked qubit initially in state 1
are

\begin{eqnarray}
\label{Pk1}
   P_1(t) &=& |a_1(t)|^2 = |U_{11}^k(t)|^2
    = \cos^2 \alpha  \nonumber \\
   P_2(t) &=& |a_2(t)|^2 = |U_{21}^k(t)|^2
    = \sin^2 \alpha \ .
\end{eqnarray}
This simple example may be extended to a series of
kicks~\cite{kaplantime}.  It is one of the few cases in which
analytic solutions may be obtained for qubits controlled by external
potentials.

\vskip 1em
{\bf Double kicks}
\vskip 1em

The simplest example of a series of arbitrary kicks is a sequence of
two kicks of strengths $\alpha_1$ and $\alpha_2$, applied at times
$t_1$ and $t_2$ respectively, i.e. $\hat V_S(t) = (\alpha_1 \delta(t
- t_1) + \alpha_2 \delta(t - t_2)) \sigma_x$.  Eq.~(\ref{uoperator})
is then easily solved in the interaction representation, namely,
\begin{eqnarray}
\label{U2a}
\hat U^{k_2,k_1} &=& \hat U^{k_2} \times \hat U^{k_1}  \nonumber  \\
 &=& \pmatrix{ \cos\alpha_2 & -ie^{-i \Delta E \, t_2} \sin\alpha_2
\cr -ie^{i \Delta E \, t_2} \sin\alpha_2 & \cos\alpha_2 } \times
\pmatrix{ \cos\alpha_1 & -ie^{-i \Delta E \, t_1} \sin\alpha_1
\cr -ie^{i \Delta E \, t_1} \sin\alpha_1 & \cos\alpha_1 }   \nonumber  \\
 &=&
\pmatrix{ U_{11} & U_{12} \cr U_{21}  & U_{22} } \ \ ,
\end{eqnarray}
where
\begin{eqnarray}
\label{Uij} U_{11} &=& \cos\alpha_1 \cos\alpha_2 - \sin\alpha_1
\sin\alpha_2\,e^{-i \Delta E \, t_-} \ , \\ \nonumber U_{21} &=& -i
e^{ i {\Delta E}\, t_+/2} (\cos\alpha_1 \sin\alpha_2 \,e^{i {\Delta
E} \,t_-/2}
    + \sin\alpha_1 \cos\alpha_2 \,e^{-i {\Delta E} \,t_-/2}) \ .
\end{eqnarray}
Here $t_- = t_2 - t_1$, and $t_+ = t_1 + t_2$.  In the limit $t_2
\to t_1$, Eq.~(\ref{U2a}) reduces to Eq.~(\ref{Usk}) with $\alpha
\to \alpha_1+\alpha_2$.  Note that $[\hat U^{k_2} , \hat U^{k_1}]
\neq 0$ so that the time ordering of the interactions is important.

The algebra for a combination of two arbitrary
kicks~\cite{shakov05}, one proportional to $\sigma_y$ and the other
proportional to $\sigma_x$, is very similar to the above.  Triple
kicks are also straightforward to solve analytically.

\subsubsection{Time ordering in a doubly kicked qubit}

In this subsection we use our analytic expressions to examine the
effect of the Dyson time ordering operator $T$ in a kicked two-state
system.  Time ordering has been considered previously in the context
of atomic collisions with charged
particles~\cite{mcbook,gm01,zhao,aarhus,bruch} and differs somewhat
from the order in which external pulses are applied, as illustrated
below.  As is intuitively evident, there is no time ordering in a
singly kicked qubit~\cite{kick} since there is only one kick.  The
simplest kicked two-state system that shows an effect due to time
ordering is the qubit kicked by two equal and opposite pulses
labeled $k$ and $-k$ separated by a time interval $t_- = t_2 - t_1$.
The evolution matrix for this system is~\cite{shakov05},
\begin{eqnarray}
\label{U-kk1}
 \hat U^{-k,k}=
\pmatrix{
 e^{-i\Delta E\,t_-/2}(\cos\frac{\Delta E}{2}t_- +i \cos 2 \alpha \sin
\frac{\Delta E}{2}t_-)    & e^{-i \Delta E \, t_+} \sin 2 \alpha
\sin \frac{\Delta E}{2} t_-       \cr
 -e^{ i \Delta E \, t_+} \sin 2 \alpha \sin \frac{\Delta E}{2} t_-  &
  e^{i\Delta E \,t_-/2}(\cos\frac{\Delta E}{2}t_- -i \cos 2 \alpha \sin
\frac{\Delta E}{2}t_-) } \ .
\end{eqnarray}
The time evolution operator $\hat U^{(0)}$ in the limit of no time
ordering, i.e. in the approximation $T \to 1$, may in principle be
generally obtained~\cite{kick} by replacing $\int_0^t \hat V(t')
\,dt' $ with $ \overline{\hat V} t $, where $ \overline{\hat V}$ is
an average (constant) value of the interaction.  In our case, it is
then straightforward to show
\begin{eqnarray}
\label{U-kk20} \hat U^{(0) -k, k }= e^{-i\bar{\hat V} t}  =
\pmatrix{
    \cos( 2 \alpha \sin \frac{\Delta E}{2} t_-)    &
  e^{-i \Delta E \, t_+} \sin(2 \alpha \sin \frac{\Delta E}{2} t_-)       \cr
 -e^{ i \Delta E \, t_+} \sin(2 \alpha \sin \frac{\Delta E}{2} t_-)  &
    \cos( 2 \alpha \sin \frac{\Delta E}{2} t_-)  } \ .
\end{eqnarray}
In this example we now have analytic expressions for the matrix
elements of both the evolution operator $\hat U^{-k ,k}$ that
contains time ordering and the evolution operator $\hat U^{(0) -k, k
}$ without time ordering.

\begin{figure}[h]
\begin{center}
\includegraphics[keepaspectratio=true,width=0.65\textwidth]{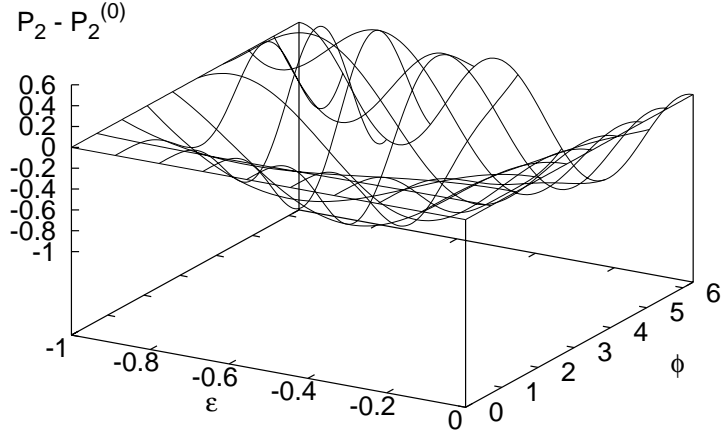}
\caption{\footnotesize{\label{f3} Difference in population transfer
probability, $P_2 - P_2^{(0)}$ vs. $\epsilon = \sin (\Delta E\,
t_-/2)$ and $\phi = 2 \alpha$.  Here $t_- = t_2 - t_1$ is the time
between the pulses and $\alpha = \int V(t')\, dt'$ is a measure of
the interaction strength.  The two-state system is kicked by a sharp
pulse of strength $\alpha$ at time $t_1$ and by an equal and
opposite pulse at time $t_2$.  The difference, $P_2 - P_2^{(0)}$, is
due to time ordering in this qubit.}} \label{figdifference}
\end{center}
\end{figure}

It may be shown analytically~\cite{shakov05} that for two kicks both
proportional to $\sigma_x$, the order of the kicks does not change
the final transfer probability $P_2$.  However, interestingly, this
does not mean that there is no effect due to time ordering in this
case.  As we show next, there is an effect due to time ordering in
this case, even though interchanging the order of the kicks has no
effect.  The effect of time ordering on the occupation probabilities
may be examined by considering the probability of population
transfer from the on state to the off state with and without time
ordering, namely from Eqs.~(\ref{U-kk1}) and (\ref{U-kk20}),

\begin{eqnarray}
\label{P2} P_2 &=& |U_{21}|^2 = |\sin 2 \alpha  \ \sin \frac{\Delta
E}{2} t_-|^2 = |\epsilon \sin \phi|^2 \ , \\ \nonumber P_2^{(0)} &=&
|U_{21}^{(0)}|^2 = |\sin( 2 \alpha \sin \frac{\Delta E}{2} t_-)|^2 =
|\sin \epsilon \phi|^2 \ \  ,
\end{eqnarray}
where $\epsilon = \sin \frac{\Delta E}{2} t_-$ and $\phi = 2 \alpha$.

The effect of time ordering is shown in Fig.~\ref{figdifference},
where $P_2 - P_2^{(0)}$ is plotted as a function of $\phi = 2
\alpha$, corresponding to the strength of the kicks, and $\epsilon =
\sin \frac{\Delta E}{2} t_-$, which varies with the time separation
of the two kicks.  The effect of time ordering disappears in our
example in the limit that either the interaction strength or the
time separation between the pulses goes to zero. For small, but
finite, values of both the interaction strength and the time
separation between the pulses, the effect of time ordering is to
reduce the transition probability from the initially occupied state
to an initially unoccupied state.  That is, in this regime time
ordering reduces the maximum transfer of population from one state
to another.  As either of these two parameters gets sufficiently
large, the effect of time ordering oscillates with increasing values
of the interaction strength or the inter-pulse separation time.
Time ordering effects are present even though $\hat U^{-k,k} = \hat
U^{k,-k}$.

\subsubsection{Time ordering vs. time reversal}

Let us now pause to examine the difference between time ordering and
time reversal in this simple, illustrative example.  Reversal of
time ordering means that, since kick strengths $\alpha_k$ and kick
times $t_k$ are both interchanged, $t_- \to -t_-$ and $\alpha \to
-\alpha$.  In this case one sees from Eqs.~(\ref{U-kk1}) and
(\ref{U-kk20}) that $\hat U^{-k ,k}$ experiences a change of phase,
while $\hat U^{(0) -k ,k}$ is invariant under change of time
ordering.  For time reversal~\cite{gw}, $t_\pm \to -t_\pm$ and,
since the initial and final states are also interchanged, $\hat{U}
\to \hat{U}^{\dag}$.  Inspection of the same equations as above
shows that $\hat U^{-k, k}$ and $\hat U^{(0) -k, k}$ are both
invariant under time reversal, as expected.  When the symmetry
between the kicks $k_1$ and $k_2$ is broken, i.e. $\alpha_2 \ne \pm
\alpha_1$, the difference between $\hat U^{k_2,k_1}=\hat U^{k_2}
\hat U^{k_1}$ and $\hat U^{k_1,k_2}=\hat U^{k_1} \hat U^{k_2}$ can
be observed~\cite{shakov05}.

\subsubsection{Numerical calculations of time ordering}

\begin{figure}[h]
\begin{center}
{\scalebox{0.5}{\includegraphics{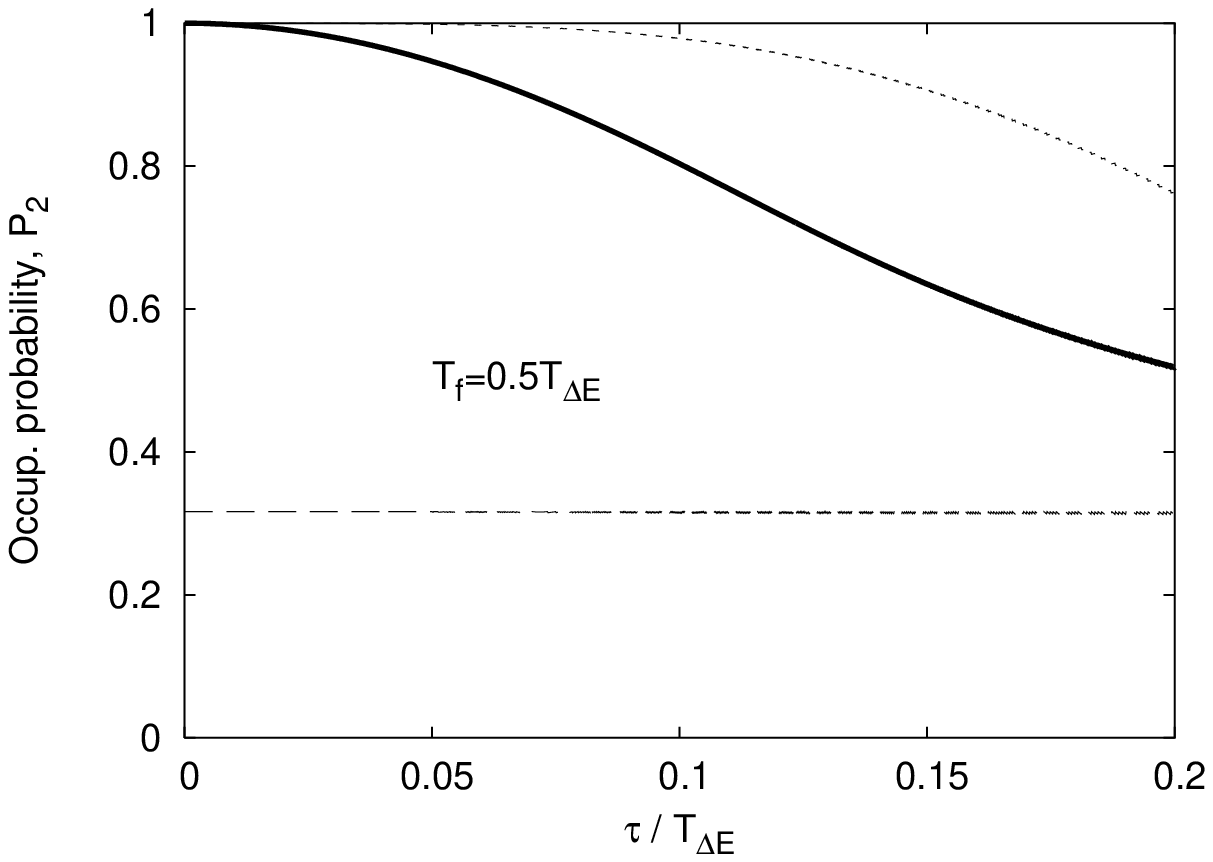}}}
{\scalebox{0.5}{\includegraphics{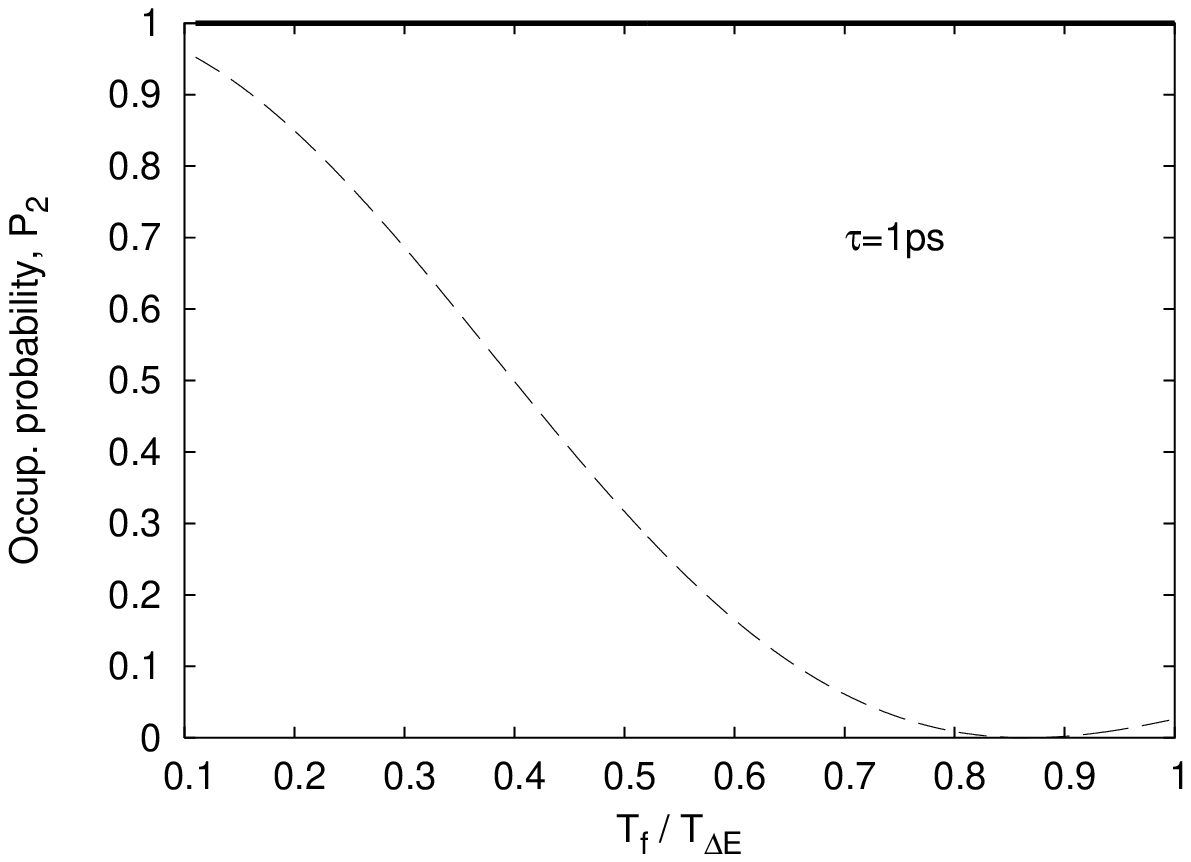}}}

{\scalebox{0.5}{\includegraphics{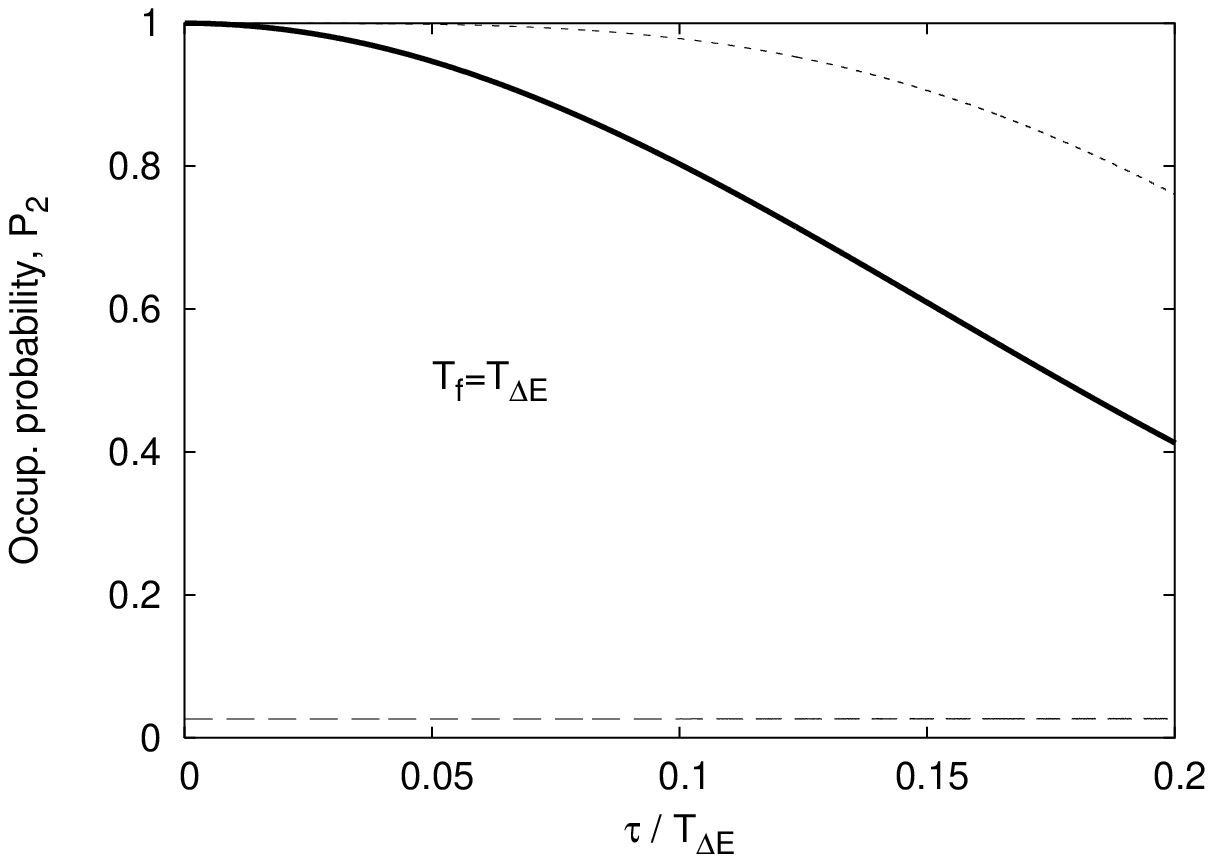}}}
{\scalebox{0.5}{\includegraphics{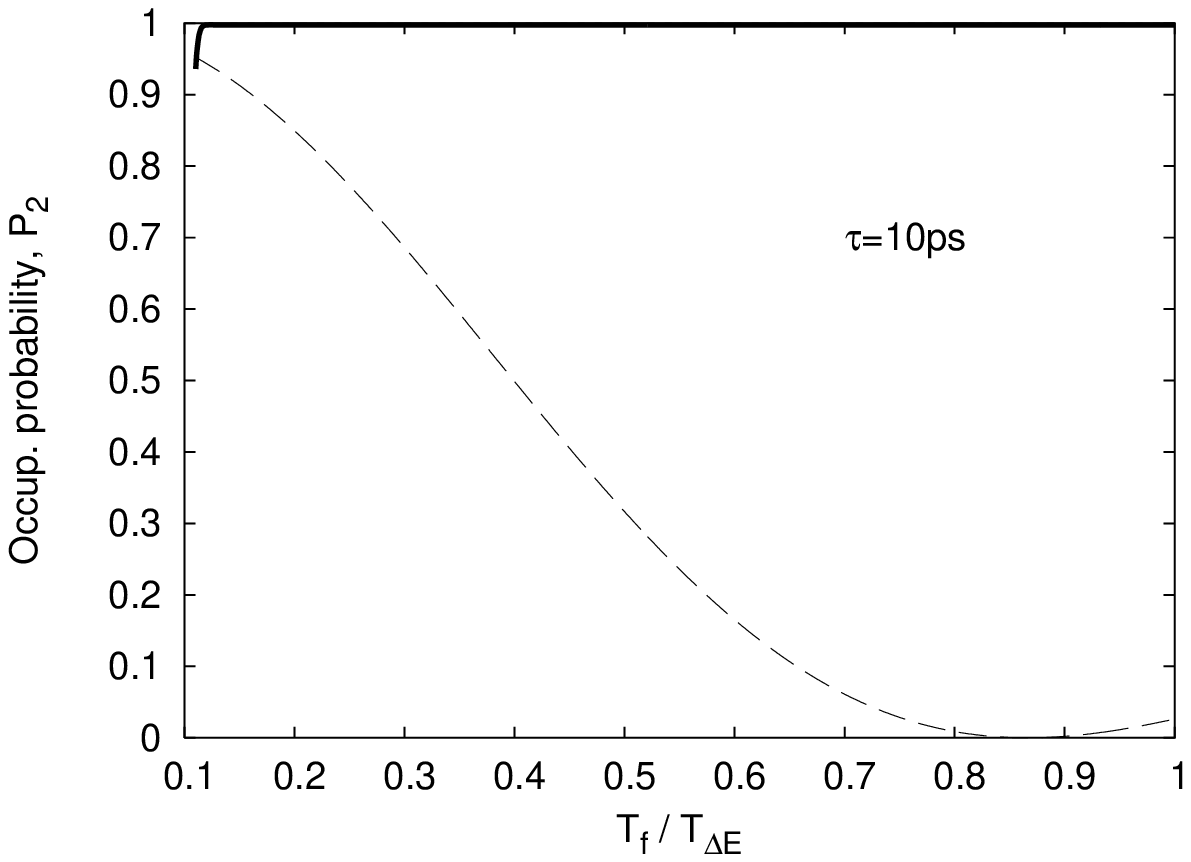}}}

{\scalebox{0.5}{\includegraphics{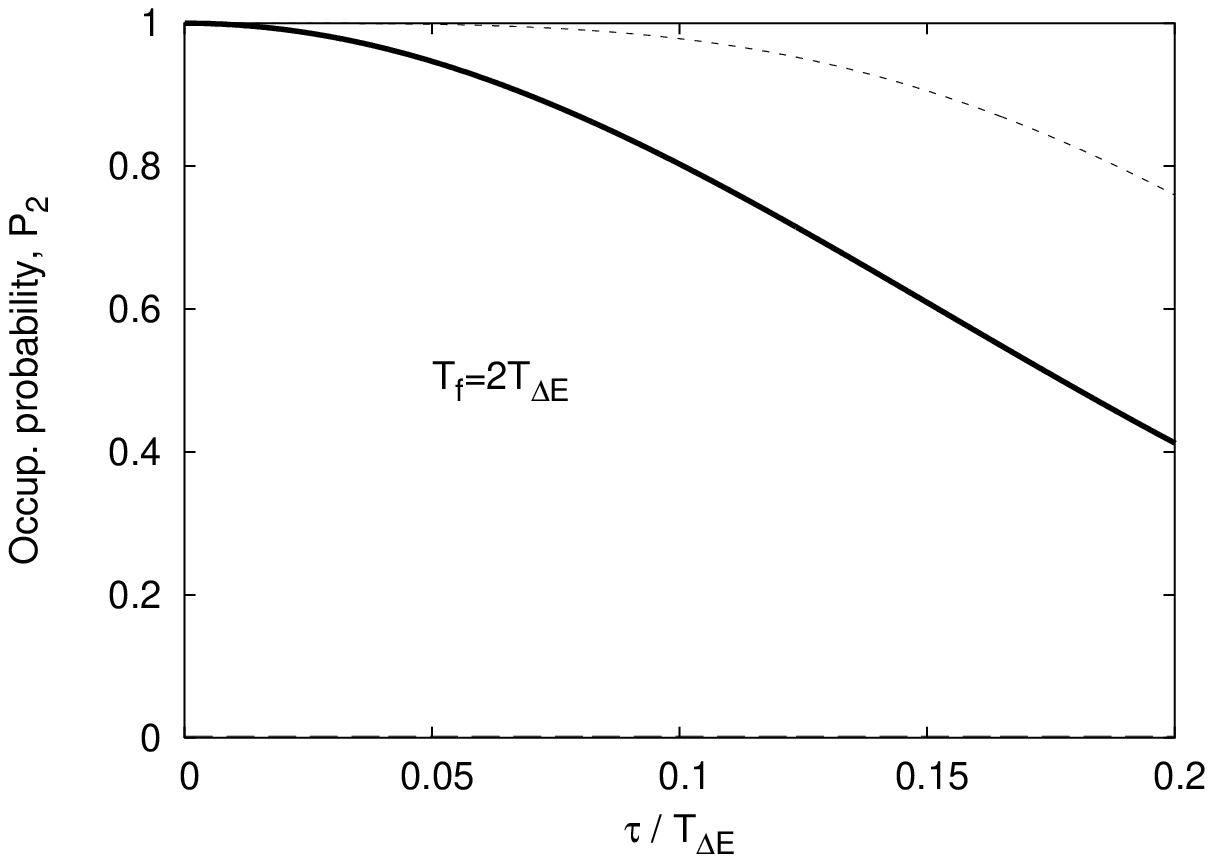}}}
{\scalebox{0.5}{\includegraphics{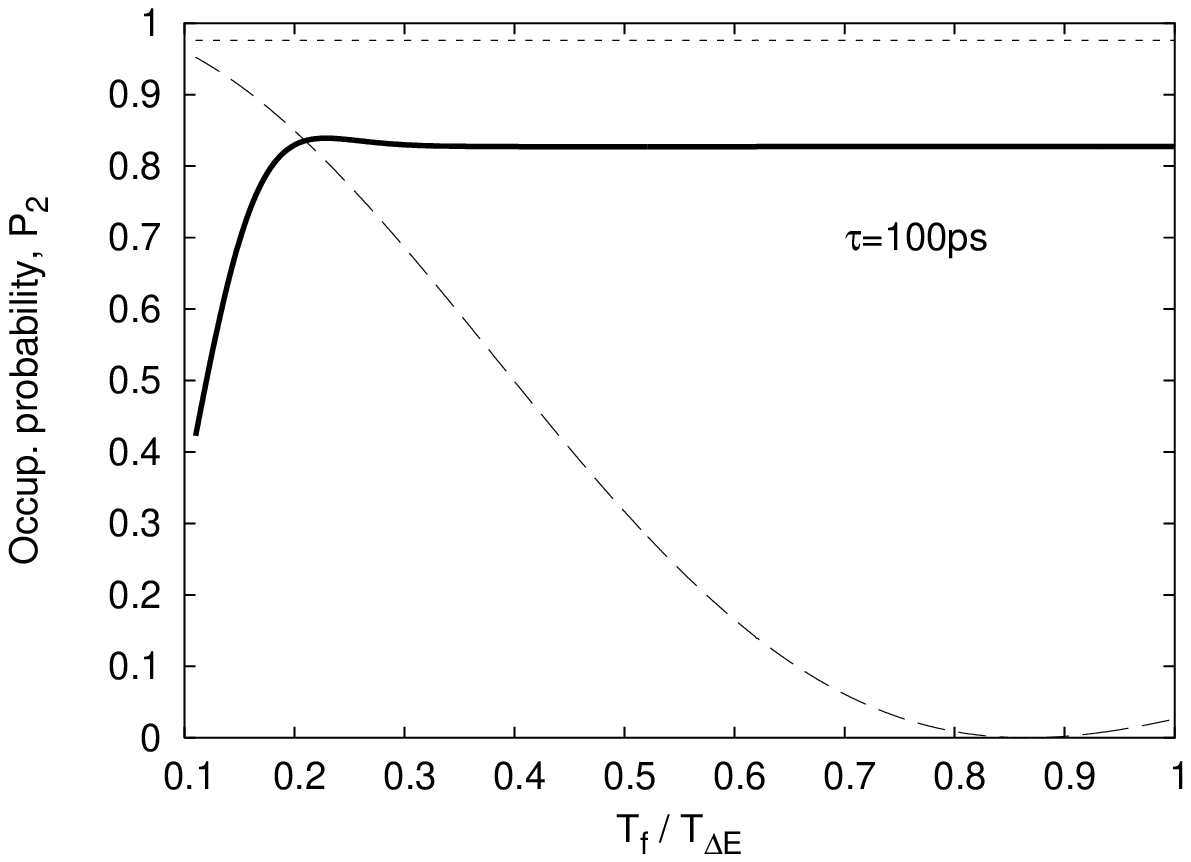}}}
\caption{\footnotesize{\label{f4} Target state probability as a
function of the pulse width $\tau$ (on the left) and of observation
time $T_f$ (on the right).  Here $T_{\Delta E}= 2 \pi  / \Delta E$
is the Rabi time for oscillations between the states, where $\Delta
E= E_{2p} - E_{2s}$.  The heavy line denotes probability including
time ordering, the dashed line denotes the probability in the
Schr\"odinger picture without time ordering, and the dotted line
represents the probability in the intermediate picture without time
ordering.  On the right, the lines begin at the midpoint of the
pulse, $T_f=t_k$.  The Schr\"{o}dinger results damp out for large
$T_f$ as explained in the text.}} \label{fignumeric}
\end{center}
\end{figure}

As an illustrative specific example, we present the results of
numerical calculations for $2s \to 2p$ transitions in atomic
hydrogen caused by a Gaussian pulse of finite width $\tau$.  The
occupation probabilities of the $2s$ and $2p$ states are evaluated
by integrating the two-state equations using a standard fourth order
Runge-Kutta method.  This enables us to verify the validity of our
analytic solutions for kicked qubits in the limit $\tau \to 0$ and
also to consider the effects of finite pulse width.  In this system,
the unperturbed level splitting is the $2s-2p$ shift, $\Delta  E=
E_{2p} - E_{2s} = 4.37 \times 10^{-6}\,{\rm eV} $.  The
corresponding time scale is the Rabi time, $T_{\Delta E}=2\pi
/\Delta E= 972$~ps, which gives the period of oscillation between
the states.  In our numerical calculations we use for convenience a
Gaussian pulse of the form $V(t) = (\alpha /\sqrt{\pi} \tau)
e^{-(t-t_k)^2/\tau^2}$.  The two-state coupled equations for the
amplitudes $a_1$ and $a_2$ of Eq.~(\ref{psi}) take the form implied
by Eq.~(\ref{genh0}),
\begin{eqnarray} \label{2s2p-1}
 i  \dot{a}_1 &=& - \frac{1}{2} \Delta E \, a_1
    + {\alpha \over \sqrt{\pi}\tau} e^{-(t-t_k)^2/\tau^2} a_2
    \nonumber \\
 i  \dot{a}_2 &=& \ \frac{1}{2} \Delta E \, a_2
    + {\alpha \over \sqrt{\pi}\tau} e^{-(t-t_k)^2/\tau^2} a_1  \ \ .
\end{eqnarray}
Here the pulse is applied at $t_k=150$ ps and we have chosen $\alpha
= \pi/2$ so that in the limit of a perfect kick all of the
population will be transferred from the $2s$ to the $2p$ state after
$t = t_k$.  In Fig.~\ref{fignumeric}, the resulting transition
probability is shown as a function of pulse width $\tau$ and as a
function of observation time $T_f$ for $T_f>t_k$.

In the Schr\"{o}dinger picture there are very large differences
between the transition probabilities with and without time ordering,
$P_2(T_f)$ and $P^{(0)}_2(T_f)$, even for an ideal kick.  This
occurs because the energy splitting $\Delta E$ is non-zero, and for
$T_f > \alpha /\Delta E = \alpha \,T_{\Delta E}/ 2\pi $, the average
potential $\bar V=\alpha/T_f$ becomes smaller than the energy
splitting $\Delta E$.  Thus, for a given pulse, the influence of the
potential necessarily decreases at large $T_f$, and any transfer
probability becomes exponentially small.  In effect, the free
propagation before and after the pulse diminishes the effect of the
pulse itself in the Schr\"{o}dinger picture, when time ordering is
removed.  This behavior contrasts with the intermediate picture
result, where $P_{I \, 2}^{(0)}(T_f)$ depends on $\Delta E \,\tau/2$
but not on $T_f$, as seen on the left side of Fig.~\ref{fignumeric}.
The contrast is also evident on the right hand side of
Fig.~\ref{fignumeric}, where, after the pulse has died off, the
value of $P_{2}^{(0)}(T_f)$ decays with increasing $T_f$, while
$P_{I \, 2}^{(0)}(T_f)$ approaches a constant.

\subsection{Networks of qubits}

To our knowledge relatively little has been understood analytically
for systems of coupled qubits.  Understanding coupling between
qubits is a well recognized challenge in the field of quantum
computing~\cite{roadmap}.  In our view, developing a realistic
analytic model for two coupled qubits could provide a useful and
instructive example in the fields of quantum computing, quantum
information and coherent control.  Unitary evolution operators
acting on a system of non-interacting qubits formally belong to the
$SU(2) \times SU(2) \times \cdots\times SU(2)$ unitary group, which
is simply the set of all local qubit operations.  This set is a
subgroup of the full $SU(2^N)$
dynamic group of N coupled qubits. The
$SU(4)$ dynamic group of two interacting qubits plays a fundamental
role in the analysis of multi-qubit dynamics since any operator
(i.e. quantum gate) from the full $SU(2^N)$ group can be factorized
as a product of $SU(4)$ two-qubit gates.  In connection with this
property, it is useful to note that any $4$-level quantum system can
be used to encode a tensor-product four-dimensional Hilbert space of
two qubits.  The specific form of such encoding is completely
determined~\cite{rau05} by fixing one point on the orbit of the
maximal $SU(2) \times SU(2)$ subgroup of $SU(4)$.  Next, the
$3$-qubit system (e.g. a carrier space for GHZ multiparticle
entangled states) can be similarly encoded in an arbitrary $8$-level
quantum system.  This requires two steps: identifying the maximal
$SU(4) \times SU(2)$ subgroup of $SU(8)$, and then adjusting the
$SU(2)\times SU(2)$ subgroup of the resulting $SU(4)$ group.

It would be useful to have a well developed theory for time
correlations between interacting qubits -- one that clarifies how
the time dependence of a field acting on one qubit impacts the time
evolution of another qubit, for example in switching.  However,
defining the independent time approximation for N-qubit systems in a
useful way is a challenge.  Although working with $n=2^N$ degenerate
states is doable in principle~\cite{rakhimov04} (setting aside the
problem of solving an $n$-th order equation for $n>4$), this does
not always give the NTO approximation, seemingly a prerequisite for
the ITA approximation needed to define time correlations.

Furthermore, in many cases sequencing of external interactions can
be problematic. For example, if $\hat V(t') = \hat V_A(t') + \hat
V_B(t')$ for two particles or qubits $A$ and $B$, and $\hat U = T
e^{-i\int (\hat V_A(t') + \hat V_B(t')) dt'}$, then the ITA is {\it
not} $\hat U_A \cdot \hat U_B$ in general.  The ITA {\it is} given
by $\hat U_A \cdot \hat U_B =\hat U_B \cdot \hat U_A$ if $[\hat
V_A(t),\hat V_B(t')] = 0$, but in that case all time correlations
vanish and the ITA is exact.  Also, it difficult for us to envision
how one may satisfy $[\hat U_A(t),\hat U_B(t')]=0$ with $[\hat
U_J(t),\hat U_J(t')] \neq 0$ for $J = A,$ $B$.   That is, in what
situations can one eliminate inter-particle time correlations while
retaining time ordering for individual particles?   If {\it all}
commutators terms vanish, then time correlation effects and time
ordering for individual particles are both absent.  Similarly, time
correlation effects disappear if all parts of a composite $\hat
V(t)$ are replaced by the time averaged value $\overline{\hat V}$,
but this again is equivalent to eliminating all time ordering, even
within single-particle evolution.


\section{Summary}

In solutions of the time dependent Schr\"{o}dinger equation there
are only two sources of time dependence, namely the Dyson time
ordering operator $T$ and the explicit time dependence of the
interaction $V(t)$. This simplifies the study of how time dependence
may influence the evolution of $N$-body quantum systems.  The
causal-like constraint of time ordering between fields acting on
different particles can cause time correlation between the
particles.  That is, the time dependence of a field acting on one
particle can influence the evolution of other particles. Correlation
is traditionally studied by defining an uncorrelated limit.  In the
case of time correlation, we have called this the independent time
approximation (ITA), and have pointed out similarities (e.g. in
Table~\ref{table1}) to the widely used and practical independent
particle approximation (IPA) that eliminates spatial correlations
between particles.  Similarly, the limit of no time ordering (NTO)
can be defined by eliminating {\it all} time ordering constraints,
for fields acting on the same particle or on different particles.
Thus, the ITA may be viewed as the NTO applied to cross terms only.

The ITA or NTO limit may be reached in several ways, but the most
general seems to be to define a mean time-averaged coupling
interaction (as is done in the IPA).  Time ordering effects have
been observed in weakly perturbed atomic collisions.  One may also
consider strongly perturbed systems of coupled qubits.  However,
relatively little has been done on this problem, which is a key
problem in quantum computing.  In this paper we considered the
effect of time ordering on strongly perturbed single qubits.  To do
this, we focused on qubits subjected to fast strong external pulses
called kicks, where useful analytic expressions for observable
transition probabilities may be obtained.  What we have found (but
not discussed here) is that the NTO may be easier to implement than
the ITA.  This means that the ITA might be computationally awkward
in general.  We have also demonstrated that the NTO (and
consequently the ITA) are dependent on the representation used.
There is evidence that the intermediate representation is preferred.
This suggests that one could find gauge dependence in specific NTO
and ITA terms if MBPT is used, and raises a
question~\cite{mcbook,uskov04} about the physical meaning of time
ordering and time correlation.

In summary, the methods we have developed in this paper are intended
to probe the nature of how time works in quantum N-body systems.
The focal point in our approach is the influence of the constraint
imposed by time ordering, which can lead to time correlation between
different parts of the system.



\begin{thebibliography}{99}
\bibitem{ita}   A.~L.~Godunov and J.~H.~McGuire, J. Phys. B {\bf 34}, L223
(2001).  Time correlation is defined as a deviation from the
independent time approximation (ITA), where $\hat U(t) = \prod_j
\hat U_j(t)$ (even if $\Psi(t_0)$ is correlated, i.e. $\Psi(t_0) =
\sum_k c_k \prod_j  \psi_j^{(k)}(r_j,t_0)$).
\bibitem{time} Spatial correlation is defined via
    ${\rm Tr} (\hat{\rho}\, \hat{r}_i^{(1)} \hat{r}_j^{(2)})
    \neq {\rm Tr} (\hat{\rho} \,\hat{r}_i^{(1)}) Tr (\hat{\rho} \,\hat{r}_j^{(2)})$
    so that $\langle\hat{r}_i^{(1)} \hat{r}_j^{(2)}\rangle \neq \langle\hat{r}_i^{(1)}\rangle \langle \hat{r}_j^{(2)}\rangle$.
    It would be interesting to explore a corresponding relation in time.
\end{thebibliography}
\end{document}